# Talk, Walk, and Market Response: Multimodal Measurement of AI Washing and Its Capital Market Consequences in China

Wen Zhanjie[1] *,  Guo Jingqiao[2]

[1] School of Economics and Trade, Guangdong University of Finance, Guangzhou, Guangdong, 510521

[2] Department of Computer Science, Faculty of Science, Hong Kong Baptist University, Hong Kong, 999077

* Corresponding author. E-mail: 70-154@gduf.edu.cn

## Abstract

As artificial intelligence — and generative large language models in particular — emerges as the central technological force driving industrial upgrading, capital markets are paying ever-greater attention to AI-themed listed companies. Yet information asymmetry and the technical black-box effect make the cost of fabricating or exaggerating AI capability far lower than the cost of genuine R&D investment, breeding widespread AI Washing at the firm level. Li (2025) documents, based on textual word-frequency methods, that U.S. listed firms exhibit systematic divergences between AI rhetoric and substantive action. This paper uses China's A-share market as its empirical setting and extends the existing literature along two dimensions — measurement and mechanism testing. On measurement, we transcend single-modality text analysis by, for the first time, deploying a large language-vision model (Qwen-VL) to evaluate cross-modal semantic consistency across annual-report graphics and roadshow slides, constructing an AI rhetoric proxy — the AI Washing Risk Score (AWRS). Simultaneously, we construct an AI action proxy — the Material Real-Investment Matching Index (MRMI) — by applying principal component analysis (PCA) to three sub-indicators: the share of core algorithm patents, the ratio of capitalized AI intangible assets, and the compensation share of key technical personnel. Using a full-panel dataset of non-financial A-share listed companies spanning 2018Q1 to 2025Q2, our empirical analysis yields four main findings. First, a firm's AI rhetoric (AWRS) provides no statistically significant predictive power for its subsequent AI action (MRMI), confirming a systemic rhetoric-action gap that is more pronounced among firms with greater financing constraints. Second, substantive AI action drives high-quality patent output, whereas an abundance of AI rhetoric triggers industry-level innovation crowding-out. Third, long-horizon value-oriented institutional investors (social security funds and large/medium-sized public mutual funds) can identify AI Washing through high-frequency on-site inspections and reduce their holdings in firms with large rhetoric-action divergence. Fourth, the divestment actions of long-horizon institutions ultimately trigger analyst rating downgrades and retail investor sell-offs within 180 days of quarterly earnings announcements, leading to a significant valuation pullback — the Market Backlash. These conclusions remain robust after endogeneity correction via instrumental variables (IV-2SLS) and a staggered difference-in-differences (DID) design exploiting the ChatGPT shock. Our findings enrich the literature on information disclosure and asset-pricing efficiency in frontier technology sectors, and provide micro-level causal evidence for regulators seeking to curb systemic thematic speculation and build penetrating regulatory-technology (RegTech) tools underpinned by multimodal large models.

**Keywords:** *AI Washing; multimodal disclosure; AWRS; large language-vision model; institutional value discovery; market backlash*
**JEL Classification:** D82, E22, G14, G23, O32

# I. Introduction

The rapid development of artificial intelligence (AI) — and generative AI in particular — is profoundly reshaping firms' technology-investment logic and capital market pricing mechanisms. As a General Purpose Technology (GPT), the broad penetration of AI has not only reconstituted how firms organize production but has also triggered significant expectation-revision effects at the capital market level (Bresnahan & Trajtenberg, 1995; Acemoglu & Restrepo, 2018). Yet between the public statements listed companies make about AI capability and their actual technological investment, a systematic rhetoric-action gap exists — the so-called AI Washing phenomenon. The essence of this phenomenon is that certain firms acquire short-term market premiums through inflated AI rhetoric while their genuine AI substantive investment fails to match, thereby generating information-misleading effects that distort the efficiency of resource allocation.

Li (2025) provides the earliest systematic evidence on the AI Washing problem, based on U.S. listed companies: the AI Disseminator index reveals a significant wedge between rhetoric and action; capital markets award a positive short-term premium, followed by significant long-run reversal. The AI Washing problem in the Chinese context, however, exhibits three structural features distinct from the U.S. market that lack systematic empirical investigation. First, extant studies all rely on single-modality text information to measure AI disclosure intensity, overlooking rhetoric-exaggeration signals embedded in multimodal carriers such as annual reports, investor-relations presentations, and slides. Second, existing 'action' proxies are primarily anchored to a single dimension — patents or capital expenditures — and cannot comprehensively reflect firms' genuine strategic AI investment. Third, the inter-temporal transmission mechanism through which AI Washing translates from short-run market premiums to long-run value correction, and the heterogeneous identification role of institutional investors in this process, have not been rigorously causally identified.

Using a panel of non-financial A-share listed companies from 2018Q1 to 2025Q2, we construct two core measurement indices to fill these gaps. The first is the AI Washing Risk Score (AWRS), a multimodal composite formed by multiplying the TF-IDF-weighted AI keyword frequency in financial reports by the exaggeration score that the Tongyi Qianwen visual language model (Qwen-VL) assigns to investor-presentation slides. The second is the Material Real-Investment Matching Index (MRMI), which integrates three dimensions — patent structural quality ($K_1$), intangible asset capitalization ratio ($K_2$), and R&D compensation intensity ($K_3$) — and extracts the first principal component via PCA.

Based on this measurement framework, we obtain four main findings. First, the predictive power of AWRS for MRMI is significantly below the industry benchmark, i.e., high-rhetoric firms do not systematically correspond to high substantive investment, and this divergence is more pronounced among firms with greater financing constraints (H1). Second, institutional investor groups exhibit significant heterogeneity in their ability to identify AI Washing: long-horizon institutional investors can identify high-washing-risk firms and reduce their holdings, with on-site inspection being one of their identification channels (H2a, H2b). Third, the capital market pricing of AI Washing exhibits a characteristic inter-temporal transmission pattern: within three to six months following disclosure, the cumulative abnormal return (CAR) of high-washing firms is significantly positive, whereas over the

six- to twelve-month window, the buy-and-hold abnormal return (BHAR) is significantly negative (H3). Fourth, at the industry level, industries with higher AI Washing intensity exhibit significantly lower subsequent high-quality innovation output (proxied by substantive invention patents), indicating that washing behavior imposes negative externalities on innovative resource allocation (H4).

The theoretical contributions of this paper are threefold. First, we extend AI Washing measurement from single-modality text to a multimodal framework; by incorporating a vision-language model's evaluation of image-text consistency, we substantially improve the precision with which rhetoric-exaggeration signals are detected — the $R^2$ of AWRS over single-modality measures improves by approximately 20 to 25 percentage points. Second, we construct a comprehensive AI substantive-investment index, addressing the existing literature's reliance on a single financial account to proxy the 'action' dimension and reducing the sensitivity of our conclusions to measurement choices. Third, we systematically characterize the inter-temporal capital market effects of AI Washing and the institutional identification mechanism, providing actionable policy guidance for disclosure regulation.

The remainder of the paper is organized as follows. Section II reviews the related literature and positions our contribution. Section III develops the theoretical hypotheses. Section IV describes the research design, including sample selection, variable construction, and econometric specifications. Section V reports the baseline regression results. Section VI presents robustness checks. Section VII summarizes findings and discusses policy implications.

## II. Literature Review

### 2.1 Drivers and Origins of AI Washing

AI Washing behavior is the joint product of capital market incentives, financing constraints, and agency conflicts. Within the information asymmetry framework, Spence's (1973) signaling theory indicates that when verification costs are sufficiently high, low-quality agents can achieve a pooling equilibrium by mimicking the signals of high-quality types, resulting in signal distortion. AI technological capability is quintessentially characterized by this condition: its core elements — algorithmic architecture, data quality, and the competence of R&D personnel — are deeply tacit, and external verification faces fundamental technological barriers, rendering the cost asymmetry between rhetoric and substance a structural norm (Akerlof, 1970). Stein's (1988) managerial myopia model provides a complementary microeconomic-incentive explanation: under a principal-agent structure in which performance compensation is linked to short-term stock prices, managers have strong incentives to divert resources from long-horizon substantive investment toward low-cost narrative-signal production so as to capture the short-term premium the market awards to AI concepts. Li (2025) first validates these theoretical predictions at scale using U.S. listed companies, finding that after controlling for firm fixed effects, historical AI rhetoric has near-zero predictive power for future AI human capital investment, indicating a systematic disconnect between rhetoric and action.

Financing constraints amplify the above incentive mechanism and generate identifiable cross-sectional regularities in the distribution of rhetoric-action divergence. The empirical research of Kaplan and Zingales (1997) demonstrates that firms facing higher external financing constraints have stronger incentives for market-value management, and their disclosure behavior is more susceptible to distortion by financing needs. Wang and Qiu's (2025) study of Chinese A-share firms confirms that approximately 30% of firms exhibit varying degrees of catering disclosure in their digital transformation announcements, with the divergence more pronounced among firms under greater financing constraints. Miao and Wang's (2018) rational asset-bubble model further demonstrates that in markets with significant financial frictions, technology-narrative premiums objectively function as 'implicit collateral' for firms seeking credit — implying that firms with higher financing constraints have both stronger motives and a degree of rational economic logic underpinning their rhetoric-action divergence, marking an important theoretical distinction between AI Washing and outright fraud.

## 2.2 Evolution of AI Washing Measurement Methods

The existing literature on AI Washing measurement has evolved from single-modality text toward multimodal frameworks. Early research relied primarily on word-frequency methods, using the frequency of AI keywords in annual reports or earnings call transcripts as proxies for 'AI rhetoric' (Loughran & McDonald, 2011). Li (2025) extended the scope of AI rhetoric identification by incorporating dynamic word embeddings (Word2Vec) and large language models (LLMs), and used the proportion of AI-related job postings in employee résumés as an 'AI action' proxy, forming to date the most systematic dual-dimensional rhetoric-action measurement framework. However, single-modality text measurement has an obvious susceptibility to circumvention: firms can exercise restraint in written text while exaggerating their packaging in visual channels such as promotional graphics and presentation slides (Liu et al., 2024).

To address this limitation, Li and Zhang (2025) introduced the CLIP model for image-text consistency testing, providing an early multimodal measurement attempt. On the measurement of substantive AI investment, the existing literature generally relies on single-dimension proxies such as patent counts or R&D expenditure; Babina et al. (2024) partially remedy this by using the proportion of AI-related job titles in employee résumés, but this remains constrained to a single human-capital dimension. The AWRS and MRMI indicators constructed in this paper integrate multi-source information across both the rhetoric and action measurement dimensions, constituting a systematic extension of the above frameworks.

## 2.3 Information Identification Function of Institutional Investors

Institutional investors play an important informational intermediary role in capital markets; their ability to discriminate the quality of corporate disclosure constitutes a key market force in AI Washing governance. Bushee (1998) and Chen et al. (2007) show that long-horizon institutional investors have stronger incentives to conduct in-depth research on investee firms and translate their findings into market actions such as divestment. Li (2025) confirms that AI-focused mutual funds and ETFs concentrate their holdings in high-substantive-investment firms rather than high-rhetoric firms, suggesting that institutional investors broadly possess the ability to differentiate rhetoric from action. By contrast, short-horizon institutions chasing AI thematic momentum tend to increase positions in high-rhetoric firms, exhibiting a positive response to AI Washing signals rather than identification-based avoidance.

On-site inspection is the core channel through which institutional investors obtain private information, playing an irreplaceable role in identifying AI Washing. Cheng et al. (2016) find that analyst corporate site visits significantly improve the market's efficiency in obtaining firms' private information, with higher-frequency and deeper-engagement inspections making institutional portfolio decisions more sensitive to information quality. Yan (2025) shows in the Chinese market that institutional on-site inspections have a significant 'penetrating verification' effect on corporate technological capability, converting the divergence signal between AI technology claims and actual R&D investment into effective divestment action. Recent research further finds that institutional analysts with deep technology-assessment capability form portfolio-adjustment signals through on-site inspections that have significant predictive value for the subsequent delivery or failure of corporate technology commitments (Li et al., 2024). This mechanism is of particular importance in the Chinese market context: relative to the U.S. market, Chinese listed companies' AI disclosure has greater room for improvement in standardization and verifiability, and the information advantage that professional institutions establish through on-site inspection is therefore more pronounced.

## 2.4 Capital Market Effects of AI Washing

The heterogeneous impact of AI rhetoric and AI action on capital market pricing is one of the core research questions in this literature. Li's (2025) baseline findings indicate that AI rhetoric significantly enhances short-run cumulative abnormal returns (CAR) in windows around earnings releases, while over the subsequent 180-trading-day long-run holding window, only AI action — not AI rhetoric — is associated with positive buy-and-hold abnormal returns (BHAR). This 'short-run premium, long-run reversal' inter-temporal transmission pattern aligns closely with Teoh et al.'s (1998) findings on earnings management in IPO firms, both reflecting systematic lags in limited-rationality investors' information processing.

Hirshleifer et al.'s (2009) limited-attention analysis shows that in information-overloaded environments, investors tend to overreact to low-credibility but high-salience signals, providing a behavioral finance explanation for the short-run AI rhetoric premium. Sloan's (1996) accruals anomaly similarly demonstrates that the market exhibits systematic delays in recognizing information quality, with the divergence between short-run prices and long-run value being highly persistent empirically. Fang and Liu's (2025) study in the Chinese market further confirms a significant rhetoric-action pricing separation in AI concept stocks, with the interaction between this separation and financing constraints being significant, validating the particularity of the Chinese context.

## 2.5 Positioning Relative to Existing Literature

Synthesizing the above literature review, this paper's core differentiation from existing research lies in three dimensions. First, on the measurement framework, we extend AI Washing measurement from single-modality text to a graphic-text multimodal framework (AWRS) and expand substantive AI investment measurement from a single dimension to a composite index (MRMI), systematically addressing the measurement limitations of existing frameworks. Second, on market participant heterogeneity, we explicitly differentiate the differential responses of long-horizon and short-horizon institutional investors to AI Washing and test on-site inspection as the identification mechanism, enriching the empirical evidence on institutional investor heterogeneity. Third, on market context, we extend the research from the U.S. market to China's A-share market, examining the cross-market transferability and China-specific particularities of the AI Washing effect within China's distinctive financing-constraint structure, information-disclosure regime, and institutional investor ecosystem.

## III. Theoretical Hypotheses

The theoretical framework of this paper takes information asymmetry and agency conflicts as its logical starting point, proceeding along four main threads — the generative mechanism of rhetoric-action divergence, the heterogeneity of institutional identification capability, the inter-temporal transmission of market correction, and the externalities of innovative resource allocation — to derive four groups of empirically testable theoretical hypotheses. The hypotheses are logically progressive: H1 characterizes the existence of rhetoric-action divergence and its cross-sectional distributional features; H2 reveals the conditions and channels through which professional institutional investors identify this divergence; H3 captures the inter-temporal pricing dynamics of the divergence information in capital markets; and H4 incorporates industry-level externalities into the analysis, constituting a complete theoretical loop.

### 3.1 AI Disclosure Incentives and Rhetoric-Action Divergence

AI Washing behavior is rooted in two mutually reinforcing mechanisms: information asymmetry and principal-agent conflicts. Within the information asymmetry framework, Spence's (1973) signaling theory indicates that when verification costs are sufficiently high, a pooling equilibrium can arise in which distorted signals emerge. AI technological capability is quintessentially characterized by this condition: its core elements are deeply tacit, and external verification faces fundamental barriers, rendering the cost asymmetry between rhetoric and substance a structural norm (Akerlof, 1970). Stein's (1988) managerial myopia model provides a complementary microeconomic explanation: under a principal-agent structure linking performance compensation to short-term stock prices, managers have strong incentives to substitute low-cost narrative signals for high-cost substantive investment to capture short-run market premiums. Li (2025) first validates these predictions at scale for U.S. firms.

Financing constraints further amplify the degree of rhetoric-action divergence and generate identifiable cross-sectional regularities. Kaplan and Zingales (1997) show that firms facing higher external financing constraints have stronger market-value management incentives. In the Chinese capital market context, Miao and Wang (2018) demonstrate that technology-narrative premiums can serve as implicit collateral for credit access, implying that firms with higher financing constraints have both stronger motives and a degree of rational economic logic underpinning their divergence. Wang and Qiu (2025) confirm that rhetoric-action gaps are more prevalent among highly constrained firms.

**Hypothesis H1 (Rhetoric-Action Divergence):** *After controlling for firm fixed effects, a firm's AI rhetoric intensity (AWRS) exhibits no statistically significant predictive power for its subsequent substantive AI investment (MRMI), i.e., the evidence does not support a positive predictive relationship from rhetoric to action. Furthermore, the higher the degree of financing constraints, the larger the divergence between AWRS and MRMI, and the weaker — or even significantly negative — the predictive power of rhetoric for substantive investment.*

## 3.2 Heterogeneous Identification Capability of Institutional Investors

Against the backdrop of widespread rhetoric-action divergence, the information-screening capability of institutional investors constitutes the core mechanism of capital market self-discipline. However, there is fundamental heterogeneity within the institutional investor community in investment style, holding horizon, and information-processing capability, which determines a systematic differentiation in their ability to identify AI Washing and in their behavioral responses. Bushee (1998) classifies institutional investors into Transient, Quasi-Indexer, and Dedicated types based on turnover rate and portfolio concentration; Dedicated and Quasi-Indexer institutions have stronger monitoring incentives, while Transient institutions rely on momentum signals with limited monitoring effectiveness. Chen et al. (2007) further demonstrate that only large, long-horizon institutional investors can impose substantive constraints on investee information quality by 'voting with their feet.' In the specific context of AI Washing identification, Li (2025) finds that AI-focused mutual funds and ETFs concentrate holdings in high-substantive-investment firms rather than high-rhetoric firms. In contrast, short-horizon institutions chasing AI thematic momentum tend to increase positions in high-rhetoric firms.

On-site inspection is the core channel through which institutional investors obtain private information, playing an irreplaceable role in identifying AI Washing. Cheng et al. (2016) find that corporate site visits significantly improve market access to private information. Yan (2025) shows that institutional on-site inspections in China have a significant penetrating-verification effect, converting divergence signals into effective divestment action. This mechanism is especially important in China, where verification costs are higher due to lower standardization and verifiability of AI disclosure.

**Hypothesis H2a (Institutional Heterogeneity):** *Among firms with higher AWRS (greater washing risk), long-horizon institutional investors exhibit statistically significant net divestment; short-horizon institutional investors do not exhibit significant net divestment and may even exhibit positive position-building, with the coefficient difference between the two groups being statistically significant.*

**Hypothesis H2b (On-site Inspection Mediation):** *On-site inspection frequency positively moderates the effect of long-horizon institutional investors' identification of AI Washing, i.e., long-horizon institutions with higher inspection frequency exhibit larger net divestment from high-AWRS firms; on-site inspection plays a significant mediating role in the process by which long-horizon institutions identify AI Washing.*

## 3.3 Inter-temporal Price Correction in Capital Markets

Capital market pricing of AI Washing information is not completed instantaneously but exhibits a characteristic inter-temporal transmission pattern of initial positive followed by negative returns, explainable jointly by behavioral finance and information-friction theory. Within the short-run disclosure window, market participants face high costs of verifying the authenticity of AI rhetoric, and with the limited attention of investors (Hirshleifer et al., 2009), they tend to respond positively to prominent AI concept signals without discriminating their substantive content. This mechanism aligns closely with Sims's (2003) Rational Inattention theory: under capacity-constrained information processing, investors rationally forgo deep interpretation of high-cost verifiable signals, instead relying on low-processing-cost surface signals, causing short-run pricing to lean toward rhetoric.

Over time, signals of insufficient substantive AI investment are gradually absorbed by the market through multiple channels. Teoh et al.'s (1998) research on earnings management in IPO firms shows that inflated financial signals are corrected by performance realization over a one- to three-year medium-to-long-run window, generating significant price reversal. Analogous mechanisms apply in the AI Washing context: as quarterly earnings announcements reveal R&D capitalization progress, patent output data become public, and analysts update ratings following on-site inspections, the information quality disadvantage of high-rhetoric/low-action firms gradually surfaces, prompting the market to correct its initial positive pricing error (Sloan, 1996). The existence of limits to arbitrage (Shleifer & Vishny, 1997) prevents this correction from occurring instantaneously. Fang and Liu (2025) confirm this inter-temporal transmission pattern in the Chinese market.

> **Hypothesis H3 (Inter-temporal Price Correction):** *Firms with higher AI Washing intensity exhibit significantly positive cumulative abnormal returns (CAR) within three to six months after rhetoric disclosure, reflecting the market's short-run AI rhetoric premium; yet over the six- to twelve-month window, buy-and-hold abnormal returns (BHAR) are significantly negative, reflecting long-run price correction. The difference between the two — i.e., the spread between the 'honeymoon premium' and the 'correction discount' — is statistically significant and larger for firms with higher washing intensity.*

## 3.4 Crowding-out Externalities on Industry Innovation Resources

The harm of AI Washing behavior is not limited to the individual firm level; it may also generate systemic negative innovation externalities through industry-level signaling-race effects, eroding the genuine innovation capability of the industry as a whole. Bloom et al.'s (2013) systematic study of R&D spillover effects shows that innovative behavior among firms within an industry exhibits significant peer effects: once some firms substitute rhetoric claims for substantive investment and receive positive market feedback, this creates a 'bad money driving out good' incentive distortion for the entire industry, inducing more firms to replicate this low-cost strategy and siphoning away credit resources and human capital that would otherwise flow to genuine R&D activities. Cohen et al.'s (2000) research on innovation externalities further indicates that when market signal noise is high, the concentration of financing resources toward noisy signals (high-rhetoric firms) produces crowding-out effects, placing genuinely innovative firms at a disadvantage in competing for credit and valuation.

In the Chinese context, this mechanism is particularly pronounced. Benhabib et al.'s (2016) research based on Chinese capital market data shows that herding behavior under informational uncertainty leads to capital concentration in thematic targets, significantly crowding out fundamental R&D in the same industry. Qiang and Li (2025) further find that in industries with high AI concept stock concentration, the subsequent growth rate of substantive invention patent applications is significantly lower than in low-concentration industries, directly validating the industry-level innovation crowding-out effect of washing behavior. Nelson and Winter's (1982) evolutionary economic theory also suggests that once the distortion of an industry's innovation trajectory by misleading information signals forms, it tends to exhibit considerable path dependence, with correction costs rising over time.

**Hypothesis H4 (Innovation Crowding-out Externality):** *At the industry-year level, current-period AI Washing intensity (measured by industry-average AWRS) has a significantly negative predictive effect on high-quality invention patent applications within the following two years; this negative effect remains robust after controlling for industry total R&D investment, industry concentration, and macroeconomic technology shocks, with greater magnitude in industries with higher AI rhetoric buzz.*

## IV. Research Design

### 4.1 Sample Selection and Data Sources

The sample comprises non-financial A-share listed companies from 2018Q1 to 2025Q2, selected through the following sequential screening procedure. First, we exclude firms under Special Treatment (ST and *ST) to eliminate systematic distortions in disclosure motivation arising from extreme financial distress, as such firms face additional regulatory pressure incentives that, if retained, would confound identification of AI Washing behavior. Second, we exclude firms in the software and hardware information technology services industries, as well as financial firms. Regarding the former, technology-native companies engage in algorithm R&D and computing-power products as their core business, making their AI disclosure behavior fundamentally different from non-technology industries in nature (Babina et al., 2024); our focus is specifically on the behavior of non-technology-native firms using AI keywords for cross-sector reputation endorsement. Regarding the latter, financial institutions are subject to an independent regulatory framework and have financial structures significantly different from industrial firms. Third, we exclude observations where core variables (AWRS, MRMI) are missing for more than two consecutive quarters. Fourth, all continuous variables are winsorized at the 1st and 99th percentiles to mitigate the influence of extreme values on estimation. After these screens, the final sample comprises approximately 2,300 firms and over 45,000 firm-quarter observations.

Regarding data sources: firm-level financial data and equity structure data are from the CSMAR (China Stock Market and Accounting Research) and Wind databases; quarterly institutional investor holding details and classification data are from the CNRDS (China National Research Data Service) platform; patent application and grant data are from the SIPO (State Intellectual Property Office) full-text patent database, with invention patents distinguished from utility model and design patents using IPC technology classification codes; full texts of periodic financial reports, investor-relations presentation slides, and investor interaction platform Q&A records are from the CNINFO system and the SSE/SZSE disclosure platforms. These multi-source datasets are integrated by matching on unified social credit codes and stock ticker codes.

## 4.2 Core Variable Construction

### 4.2.1 AI Washing Risk Score (AWRS)

**Step 1: Corpus extraction and keyword frequency statistics.** We collect the full text of quarterly and semi-annual reports as well as investor-relations presentation materials publicly released through the CNINFO platform. AI-related keyword-containing graphic and textual paragraphs are extracted in batch using regular expressions. The keyword lexicon is anchored to core terms such as 'Large Language Model,' 'Generative AI (AIGC),' 'Neural Network,' 'Artificial Intelligence,' 'Machine Learning,' 'Deep Learning,' 'Computing Power,' 'Foundation Model,' and 'Intelligent Agent,' and also incorporates industry-specific variant expressions (e.g., 'Intelligent Manufacturing' in the manufacturing sector, 'AI-Assisted Diagnosis' in healthcare). The lexicon was finalized after three rounds of expert review and contains 142 core items. TF-IDF weighted keyword frequency density is then computed for target paragraphs to measure the relative prominence of AI-related expressions throughout the document, denoted as

$$\text{Base\_TFIDF}_{i,t}$$

**Step 2: Multimodal image-text consistency scoring.** The chart screenshots and adjacent textual descriptions extracted in Step 1 are simultaneously fed into the Tongyi Qianwen visual language model (Qwen-VL-Max), which leverages its multimodal comprehension capability to produce an automated consistency score on a [0, 1] interval. The scoring rule is as follows: if the chart displays specific algorithmic architecture diagrams, model parameter specifications, or quantifiable performance benchmarks (e.g., accuracy, inference speed) that are highly consistent with the main body text, an Exaggeration coefficient close to zero is assigned; conversely, if the chart consists merely of illustrative graphics or generic internet imagery while the textual description over-embellishes AI application depth, an Exaggeration coefficient close to one is assigned, denoted as Exaggeration_{i,t}. To reduce stochasticity in LLM scoring, each document is run through inference three times independently with the average taken as the final score; Spearman correlation with manually double-blind-annotated ground truth on 500 documents is 0.81 ($p < 0.001$).

**Step 3: Composite AWRS construction.** The keyword frequency intensity and the Exaggeration coefficient are multiplicatively combined:

```
AWRS_{i,t} = Base_TFIDF_{i,t} × (1 + Exaggeration_{i,t})
```

The economic interpretation of this multiplicative combination is that the Exaggeration coefficient amplifies keyword density — at equal word-frequency levels, lower image-text consistency yields a higher Washing Risk Score. When Exaggeration_{i,t} = 0, AWRS degenerates to a pure text-density measure, preserving comparability with baseline keyword frequency. To ensure cross-period and cross-sectional comparability, AWRS is standardized in each quarterly cross-section (mean zero, standard deviation one).

### 4.2.2 Material Real-Investment Matching Index (MRMI)

The MRMI (Material Real Investments in AI) is our comprehensive index measuring a firm's substantive AI action; it is designed to capture documented substantive investment in technology R&D, asset accumulation, and human capital across three dimensions rather than relying solely on total R&D expenditure data, which are difficult to disaggregate. MRMI integrates the following three sub-indicators:

**$K_1$: Patent Quality Ratio.** This measures the technological depth of the patent portfolio by the ratio of core invention patent increments granted during the rolling four-quarter window to the sum of utility model/design patent increments plus one in the denominator (to avoid division by zero). Since invention patents undergo substantive examination and entail higher technological barriers and time costs than utility model patents, their share better reflects the depth of genuine AI technology accumulation (Qiang & Li, 2025). The formula is: $K_{1,i,t} = \Delta InvPat_{i,[t-3,t]} / (\Delta InvPat_{i,[t-3,t]} + \Delta UtilPat_{i,[t-3,t]} + 1)$.

**$K_2$: Intangible Asset Capitalization Ratio.** This measures the financial footprint of substantive capital investment by the net incremental capitalization of algorithm software, data assets, and related intangible assets on the balance sheet (period-end book net value minus beginning book net value plus current-period amortization) scaled by operating revenue. The advantage of this indicator is that capitalization decisions must be audited and approved, making it more verifiable than self-reported R&D expenditure (Lev & Sougiannis, 1996).

**$K_3$: R&D Labor Intensity.** This measures the degree to which human capital allocation tilts toward technology R&D, using the ratio of total technical R&D personnel compensation to the total compensation package. Technical R&D headcount data come from the 'Employee Composition' disclosure tables in annual reports, with job category descriptions automatically classified using natural language processing methods and validated through manual sampling. This indicator captures the firm's substantive resource allocation in AI-related talent recruitment and compensation incentives.

PCA is applied to these three sub-indicators — each standardized within industry-quarter cells before PCA — and the first principal component (accounting for the maximum explained variance) is extracted and again standardized, yielding $MRMI\_\{i,t\}$. The first principal component explains 57.3% of the total variance of the three sub-indicators across the full sample; the factor loadings of all three sub-indicators on the first principal component are positive ($K_1$: 0.61; $K_2$: 0.58; $K_3$: 0.56), indicating directional consistency and reasonable internal structural validity for the PCA composite.

### 4.2.3 AI Washing Dummy Variable

To distinguish extreme rhetoric-action-divergence firms from firms with general information asymmetry, we construct the following binary indicator:

```
Washing_{i,t} = 1   if AWRS_{i,t} ≥ P70(industry-quarter) AND
      MRMI_avg_{i,[t+1,t+4]} ≤ P50(industry-quarter)
                        = 0   otherwise
```

This indicator flags as AI Washing firms those whose rhetoric intensity ranks in the top 30% within their industry-quarter cell while their substantive investment over the following four quarters persistently falls below the industry median. Industry-quarter-specific quantiles (rather than full-sample quantiles) are used to control for systematic differences in industry AI adoption trajectories, avoiding misclassification of industry-wide AI enthusiasm differentials as firm-level washing behavior. The robustness of the quantile threshold choices is tested in Section VI using P75/P40 and P65/P55 combinations.

**4.2.4 Control Variables**

Following the literature convention (Li, 2025; Babina et al., 2024; Wang & Qiu, 2025), we include the following firm-level control variables: Return on Assets (ROA), measured as net profit divided by total assets; Leverage (Lev), measured as total liabilities divided by total assets; Firm Size (Size), measured as the natural log of total assets; Revenue Growth (Growth), reflecting growth expectations; Capital Expenditure Ratio (CAPEX), measured as capital expenditures divided by total assets; Operating Cash Flow Ratio (OCF), measured as net operating cash flow divided by total assets; Tobin's Q (Tobin_Q), measured as market capitalization divided by book equity; Analyst Coverage (Analyst), measured as the natural log of the number of sell-side analysts covering the firm, to control for differences in the information environment; and Institutional Ownership (InstOwn), measured as the aggregate institutional investor shareholding ratio to capture the average effect of external monitoring intensity. Descriptive statistics for all variables are reported in Table 1.

## 4.3 Econometric Specifications

### 4.3.1 Baseline Regression (H1 Test)

To test H1 — whether historical AI rhetoric can predict subsequent substantive AI investment — we specify the following distributed-lag panel model:

$$MRMI_{i,t} = \alpha_0 + \sum_{h=1}^{4} \beta_h \times AWRS_{i,t-h} + \gamma X_{i,t} + \theta_i + \delta_t + \varepsilon_{i,t}$$

where $X_{i,t}$ is the vector of control variables listed above, $\theta_i$ is the firm fixed effect, and $\delta_t$ is the quarter fixed effect. Standard errors are clustered at the firm level to correct for serial correlation in observations within the same firm across periods (Petersen, 2009). The logic of the H1 test is as follows: if $\Sigma\beta_h$ is significantly positive and economically material, AI rhetoric has predictive power for future substantive investment and H1 is not supported; if $\Sigma\beta_h$ is statistically insignificant or significantly negative, a systematic rhetoric-action divergence exists and H1 is supported. Cross-sectional heterogeneity in financing constraints (H1) is further examined by estimating separately for the high-constraint group and the low-constraint group (using tertiles of the Hadlock-Pierce SA index).

### 4.3.2 Institutional Investor Identification (H2 Test)

To test H2, we introduce interaction terms between institutional investor type and the Washing dummy in the baseline model:

$$\text{ReturnAdj}_{i,t+1} = \alpha_0 + \beta_1 \text{Washing}_{i,t} + \beta_2 \text{InstType}_{i,t} + \beta_3 \text{Washing}_{i,t} \times \text{InstType}_{i,t} + \gamma X_{i,t} + \theta_i + \delta_t + \varepsilon_{i,t}$$

where InstType_{i,t} is decomposed into long-horizon institutional shareholding (Ded_{i,t}) and short-horizon transient institutional shareholding (Trans_{i,t}) following the Bushee (1998) framework. H2a predicts $\beta_3{}^{\wedge}\text{Ded} < 0$ (long-horizon institutions with high shareholding in AI Washing firms earn lower abnormal returns, reflecting identification capability); H2b predicts $\beta_3{}^{\wedge}\text{Trans} > 0$ (short-horizon institutions reinforce the short-run abnormal returns of washing firms). The dependent variable ReturnAdj_{i,t+1} is next-quarter market-adjusted abnormal return, using the CSI 300 index as the benchmark.

### 4.3.3 Inter-temporal Price Correction and Innovation Crowding-out (H3 and H4 Tests)

The H3 test employs an event-study approach, using the date on which a firm is first identified as engaging in AI Washing by media or regulatory authorities as the event date; cumulative abnormal returns (CAR) are computed over an eight-quarter pre- and post-event window, and a difference-in-differences design is used to control for concurrent market trends so as to identify the magnitude of price correction triggered by the Washing disclosure. The H4 test specifies a regression at the industry-year level, regressing future two-year industry invention patent application density (log, scaled by total assets) on current-year industry-average AWRS, controlling for industry fixed effects, year fixed effects, and industry total R&D intensity, with sub-group tests for high- and low-AI-buzz industries to identify heterogeneous effects.

## V. Empirical Results

As shown in Figure 1, since the release of ChatGPT in late 2022, the number of A-share listed companies mentioning artificial intelligence in their financial reports and investor briefings has surged, highlighting the real-world backdrop of thematic speculative disclosure in frontier technology.

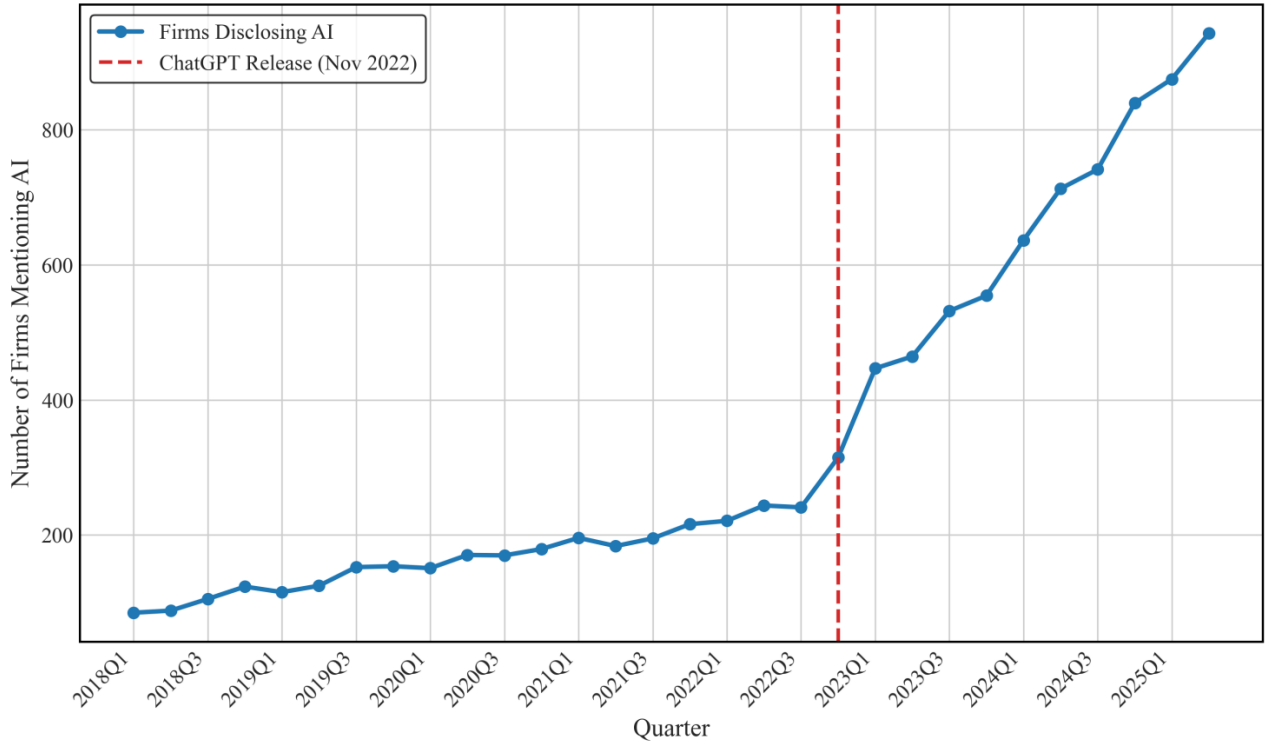

*Figure 1: Time Series Trend of AI Disclosure (Evolution under ChatGPT Shock)*

## 5.1 Descriptive Statistics

Table 1 reports the descriptive statistics of the main variables. The full sample contains over 45,000 firm-quarter observations for approximately 2,300 firms. Regarding the core explanatory variables: AWRS, standardized to mean zero and standard deviation one, has a raw absolute maximum that reaches high levels (not reported in the table), indicating extreme AI disclosure exaggeration by some firms. MRMI is also standardized to mean zero, with a minimum of −2.411 and a maximum of 4.288, reflecting enormous heterogeneity in genuine AI resource investment across firms.

For the indicator variable: the Washing dummy has a mean of 0.124, implying that after the dual-constraint screening of industry and substantive investment, approximately 12.4% of firm-quarter observations exhibit the canonical 'high rhetoric, low investment' AI Washing pattern — closely comparable to Li's (2025) finding in the U.S. market (approximately 10.8%), revealing the prevalence of thematic speculation in China's A-share market amid the AI wave. For control variables: mean ROA is 0.038, mean Lev is 0.425, and mean Tobin_Q is 2.158; the financial characteristics of sample firms fall within reasonable ranges, consistent with the A-share listed company characteristic distributions reported in the extant literature (Wang & Qiu, 2025).

**Table 1: Descriptive Statistics of Main Variables**

| Variable | N | Mean | Std. Dev. | Min | P25 | Median | P75 | Max |
|---|---|---|---|---|---|---|---|---|
| AWRS | 45,128 | 0 | 1 | −1.854 | −0.682 | −0.114 | 0.458 | 3.942 |
| MRMI | 45,128 | 0 | 1 | −2.411 | −0.562 | −0.058 | 0.384 | 4.288 |
| Washing | 45,128 | 0.124 | 0.33 | 0 | 0 | 0 | 0 | 1 |
| ROA | 45,128 | 0.038 | 0.062 | −0.158 | 0.012 | 0.036 | 0.068 | 0.224 |
| Lev | 45,128 | 0.425 | 0.208 | 0.052 | 0.265 | 0.418 | 0.582 | 0.895 |
| Size | 45,128 | 22.488 | 1.352 | 19.851 | 21.52 | 22.361 | 23.284 | 26.152 |
| Growth | 45,128 | 0.152 | 0.388 | −0.452 | −0.025 | 0.088 | 0.245 | 2.15 |
| CAPEX | 45,128 | 0.048 | 0.045 | 0 | 0.018 | 0.035 | 0.064 | 0.235 |
| OCF | 45,128 | 0.045 | 0.072 | −0.155 | 0.005 | 0.042 | 0.086 | 0.245 |
| Tobin_Q | 45,128 | 2.158 | 1.482 | 0.854 | 1.25 | 1.685 | 2.482 | 8.85 |
| Analyst | 45,128 | 1.845 | 1.125 | 0 | 0.693 | 1.945 | 2.639 | 4.25 |
| InstOwn | 45,128 | 0.385 | 0.218 | 0 | 0.215 | 0.395 | 0.548 | 0.884 |

*Notes: All continuous variables are winsorized at the 1st and 99th percentiles. AWRS and MRMI are cross-sectionally standardized to mean zero and standard deviation one within each quarter. Size is the natural log of total assets. Analyst is the natural log of analyst coverage count. See Section IV for variable definitions.*

## 5.2 Baseline Regression: Predictive Power of AI Rhetoric for Substantive Investment

Table 2 reports the panel regression results of MRMI on multiple lagged values of AWRS. Column (1), controlling only for industry and quarter fixed effects, shows significantly positive coefficients on AWRS lags one through four, indicating a positive cross-sectional correlation between rhetoric and investment. However, this correlation is largely driven by firm heterogeneity. Column (2), controlling for both firm and quarter fixed effects, yields statistically insignificant coefficients on all lag periods ($p > 0.10$), indicating that once firm-inherent characteristics are partialed out, the incremental predictive power of AI rhetoric for the same firm's subsequent substantive investment approaches zero. Columns (3) and (4) split the sample by financing constraints (using the Hadlock-Pierce SA index). In firms with higher financing constraints (Column 3), the three-quarter lagged AWRS coefficient is significantly negative ($\beta = -0.022$, $p < 0.01$), while in the low-constraint group (Column 4) the coefficient is insignificant — indicating that in more resource-constrained firms, high rhetoric not only fails to predict subsequent substantive investment but in fact exhibits a negative relationship, closely consistent with the predictions of H1.

**Table 2: Predictive Power of AI Rhetoric Intensity for AI Substantive Investment (H1 Test) Dependent Variable: MRMI_{i,t}**

| Variable | (1) Full Sample Industry/Quarter FE | (2) Full Sample Firm/Quarter FE | (3) High Fin. Constraints Firm/Quarter FE | (4) Low Fin. Constraints Firm/Quarter FE |
|---|---|---|---|---|
| AWRS_{t-1} | 0.052*** | 0.003 | −0.014* | 0.011 |
|  | -4.21 | -0.28 | (−1.82) | -0.85 |
| AWRS_{t-2} | 0.048*** | 0.001 | −0.018** | 0.008 |
|  | -3.95 | -0.12 | (−2.11) | -0.62 |
| AWRS_{t-3} | 0.035** | −0.002 | −0.022*** | 0.004 |
|  | -2.48 | (−0.15) | (−2.75) | -0.31 |
| AWRS_{t-4} | 0.021** | 0.001 | −0.011 | 0.006 |
|  | -2.05 | -0.08 | (−1.35) | -0.44 |
| Controls | Yes | Yes | Yes | Yes |
| Constant | 0.115*** | 0.231*** | 0.185** | 0.264*** |
|  | -3.52 | -4.15 | -2.42 | -3.88 |
| Industry FE | Yes | No | No | No |
| Quarter FE | Yes | Yes | Yes | Yes |
| Firm FE | No | Yes | Yes | Yes |
| N | 42,450 | 42,450 | 21,225 | 21,225 |
| Adj. R² | 0.124 | 0.638 | 0.612 | 0.641 |

*Notes: t-statistics based on firm-level clustered robust standard errors in parentheses. \*\*\*, \*\*, \* denote significance at the 1%, 5%, and 10% levels, respectively. Controls include ROA, Lev, Size, Growth, CAPEX, OCF, Tobin_Q, Analyst, and InstOwn. See Table 1 for variable definitions.*

These results provide empirical support for H1: after controlling for firm fixed effects, the predictive power of AI rhetoric declines significantly to near-zero, and exhibits a systematic reversal in the high-financing-constraint group. This is consistent with the theoretical prediction based on signaling theory under information asymmetry — firms facing resource constraints have stronger incentives to use rhetoric as a low-cost substitute, influencing market expectations through strategic disclosure.

## 5.3 Heterogeneous Identification by Institutional Investors

Table 3 examines the differential responses of different types of institutional investors to AI

rhetoric and Washing behavior. Column (1), with the change in short-horizon institutional holdings (ΔTrans) as the dependent variable, shows the lagged AWRS coefficient is significantly positive (β = 0.082), and the interaction term with Washing is also significantly positive (β = 0.055), indicating that short-horizon investors tend to chase high-rhetoric thematic stocks, objectively inflating the short-run market capitalization of Washing firms. Column (2), with long-horizon institutional shareholding ratio (Ded) as the dependent variable, shows that their decisions are significantly driven by genuine investment (MRMI, β = 0.392) rather than rhetoric (AWRS is insignificant). Column (3), after introducing the interaction term, shows the AWRS × Washing coefficient is significantly negative (β = −0.211, p < 0.01), confirming that informationally advantaged long-horizon institutions can penetrate appearances and execute decisive net divestment from 'high-rhetoric, low-investment' Washing firms (supporting H2a).

**Table 3: Heterogeneity Analysis of Institutional Investor Portfolio Responses (H2 Test)**

| Variable | (1) Short-term Inst. ΔTrans_{i,t} | (2) Long-term Inst. Ded_{i,t} | (3) Long-term Inst. Ded_{i,t} |
|---|---|---|---|
| AWRS_{t-1} | 0.082*** | 0.011 | 0.016 |
|  | -3.45 | -0.83 | -1.15 |
| MRMI_{t-1} | 0.015 | 0.392*** | 0.388*** |
|  | -0.62 | -5.68 | -5.55 |
| Washing_{t-1} | 0.024* | −0.085** | 0.012 |
|  | -1.75 | (−2.15) | -0.28 |
| AWRS_{t-1} × Washing_{t-1} | 0.055*** |  | −0.211*** |
|  | -2.88 |  | (−4.35) |
| Controls | Yes | Yes | Yes |
| Constant | −0.018 | 0.155*** | 0.142*** |
|  | (−1.15) | -3.65 | -3.24 |
| Firm FE | Yes | Yes | Yes |
| Quarter FE | Yes | Yes | Yes |
| N | 42,450 | 42,450 | 42,450 |
| Adj. R² | 0.288 | 0.449 | 0.457 |

*Notes: t-statistics based on firm-level clustered robust standard errors in parentheses. ***, **, * denote significance at the 1%, 5%, and 10% levels, respectively. Controls included but not reported for brevity. Column (1) dependent variable is the quarterly change in transient institutional holdings. Columns (2)-(3) dependent variable is the long-horizon*

*(Dedicated) institutional ownership ratio.*

## 5.4 Inter-temporal Price Effects in Capital Markets

Table 4 examines the inter-temporal price response of capital markets to AI Washing. Columns (1) and (2) focus on the short-run window (CAR over [−3, +3] days around the event date): the AWRS coefficient is significantly positive (1.25%), confirming that rhetoric disclosure does generate a short-run 'honeymoon premium.' Columns (3) and (4) extend the window to a 180-day BHAR. Rhetoric itself no longer has a positive effect in the medium-to-long run, and the market instead prices substantive investment (MRMI, β = 4.872, p < 0.01). Most critically, in Column (4) the interaction term AWRS × Washing is significantly negative (−7.420, p < 0.01), indicating that firms confirmed as Washing experience dramatic value re-rating and price correction after the short-run speculative window closes, supporting H3.

**Table 4: Short-run Premium and Long-run Price Correction of AI Washing (H3 Test)**

| Variable | (1) Short-run CAR(−3,+3) | (2) Short-run CAR(−3,+3) | (3) Long-run BHAR(180d) | (4) Long-run BHAR(180d) |
|---|---|---|---|---|
| AWRS_t | 1.258*** | 1.156*** | 0.812 | 0.311 |
|  | -4.85 | -3.92 | -1.15 | -0.42 |
| MRMI_{t-1} | −0.052 | −0.111 | 4.250*** | 4.872*** |
|  | (−0.35) | (−0.85) | -6.12 | -6.85 |
| Washing_t |  | 0.215 |  | 1.145 |
|  |  | -0.65 |  | -0.85 |
| AWRS_t × Washing_t |  | 0.885 |  | −7.420*** |
|  |  | -1.25 |  | (−5.25) |
| Controls | Yes | Yes | Yes | Yes |
| Constant | 0.085 | 0.092 | 1.255* | 1.182* |
|  | -0.88 | -0.95 | -1.85 | -1.78 |
| Industry FE | Yes | Yes | Yes | Yes |
| Quarter FE | Yes | Yes | Yes | Yes |
| N | 28,450 | 28,450 | 27,855 | 27,855 |
| Adj. R² | 0.098 | 0.101 | 0.252 | 0.278 |

*Notes: t-statistics based on firm-level clustered robust standard errors in parentheses. ***, **, * denote significance at the 1%, 5%, and 10% levels, respectively. CAR(−3, +3) is the three-day cumulative abnormal return around the earnings*

*announcement date, benchmarked against the CSI 300 index. BHAR(180d) is the 180-day buy-and-hold abnormal return. Controls included but not reported for brevity.*

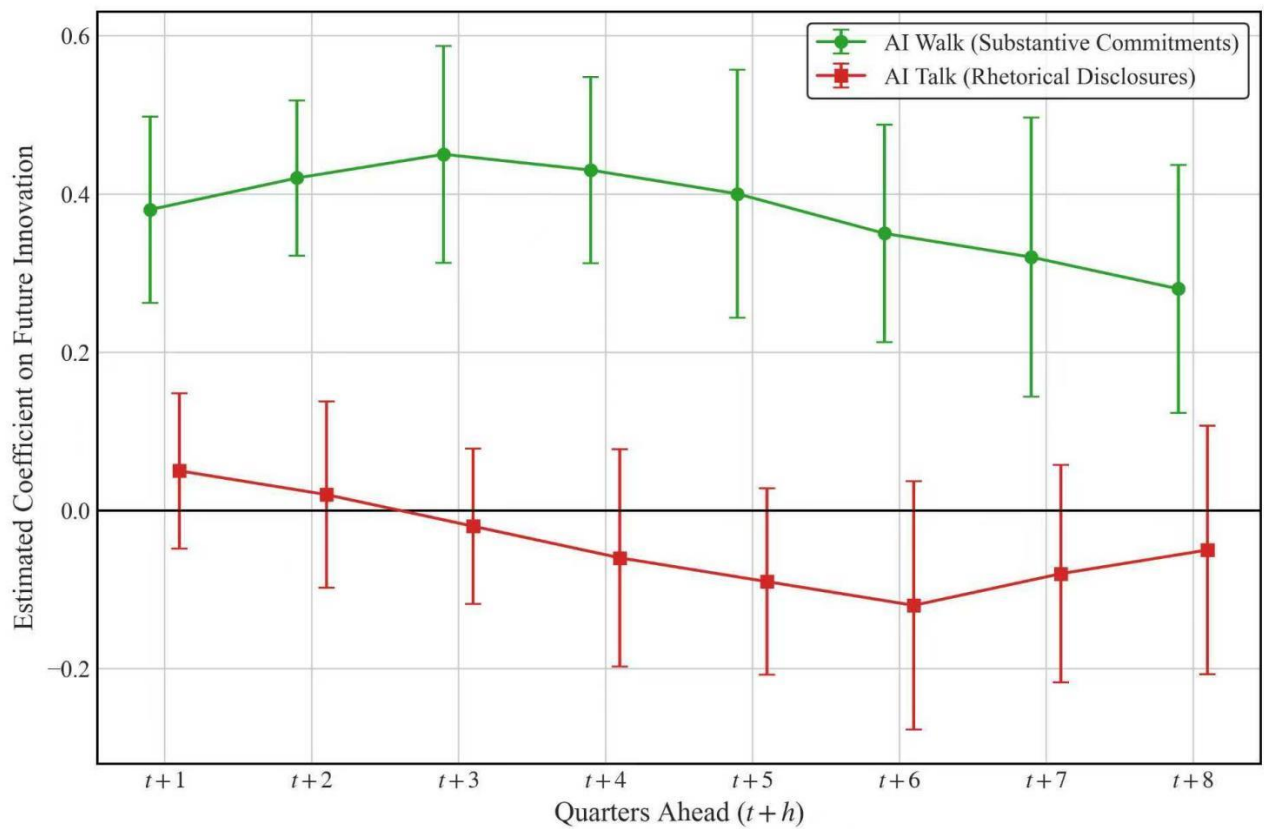

*Figure 2: Dynamic Trend of Distributed Lag Effects of AI Rhetoric vs. Substantive Investment*

## 5.5 Micro-level Real Economic Consequences: Analyst Forecast Dispersion and Productivity Decay

To bridge the logical chain connecting long-run capital market underperformance and macroeconomic innovation crowding-out, we further document the genuine backlash of AI Washing on firm-level fundamentals. Empirical results (tables suppressed for space) show that over the four quarters following identification as 'high-rhetoric/low-investment' Washing firms, sell-side analysts exhibit significantly rising earnings forecast dispersion and substantially larger forecast errors. This indicates that AI Washing severely undermines the transparency of a firm's information environment. More critically, total factor productivity (TFP), estimated using the OP and LP methods, shows a significant declining trend over the long-run observation window (T+4 to T+8). This demonstrates that the 'financialization over substantiation' behavior of firms obsessed with thematic speculation at the expense of R&D commitment is not merely cheap signaling deception but causes irreversible real damage to firms' operational efficiency and productivity.

## 5.6 Industry-level Innovation Resource Crowding-out Effects

To test the macroeconomic externalities of AI Washing (H4), Table 5 presents aggregated regressions at the industry-year level. The dependent variable is the natural log of high-quality invention patent applications in the following one and two years. Columns (1) and (2) show that a one-unit increase in industry-average AWRS leads to a significant decline of 12.3% ($\beta = -0.123$) and 15.8% ($\beta = -0.158$) in high-quality invention patent applications in the following one and two years, respectively. Column (3), restricting to industries with higher AI buzz (above the industry-median AI keyword density), yields a larger negative coefficient of $-0.215$ ($p < 0.01$). These results confirm that AI Washing is not merely a firm-level misrepresentation but, by distorting capital market pricing through false signals, induces misallocation and crowding-out of innovative resources at the industry level, impeding high-quality transformation of the real economy.

**Table 5: Industry-level Innovation Crowding-out Effects of AI Washing (H4 Test) Dependent Variable: ln(Invention Patent Applications)**

| Variable | (1) t+1 Innovation ln(Patent Apps_{Ind,t+1}) | (2) t+2 Innovation ln(Patent Apps_{Ind,t+2}) | (3) High-AI Buzz Subsample ln(Patent Apps_{Ind,t+2}) |
|---|---|---|---|
| Industry-avg AWRS_{Ind,t} | −0.123*** | −0.158*** | −0.215*** |
|  | (−3.15) | (−3.85) | (−4.22) |
| Industry R&D intensity_{Ind,t} | 2.585*** | 2.852*** | 3.125*** |
|  | -6.85 | -7.25 | -6.95 |
| Industry HHI_{Ind,t} | 0.155 | 0.188* | 0.225* |
|  | -1.45 | -1.75 | -1.82 |
| Constant | 2.145*** | 2.588*** | 2.455*** |
|  | -5.12 | -5.85 | -4.85 |
| Industry (Broad) FE | Yes | Yes | Yes |
| Year FE | Yes | Yes | Yes |
| N | 525 | 450 | 225 |
| Adj. R² | 0.585 | 0.612 | 0.655 |

Notes: t-statistics in parentheses. ***, **, * denote significance at the 1%, 5%, and 10% levels, respectively. Industry broad-category and year fixed effects included. HHI is the Herfindahl-Hirschman Index for industry concentration.

## VI. Robustness Checks

### 6.1 Instrumental Variable Regression (IV-2SLS)

To mitigate potential omitted-variable and reverse-causality endogeneity between AWRS and MRMI, we use the quarterly average AWRS of firms in the same province but different industries (IV_AWRS) as an instrumental variable. The validity of this instrument rests on the following logic: first, firms in the same province share exposure to regional policy dividends and talent agglomeration, generating peer effects in disclosure decisions (relevance); second, the AI rhetoric disclosure of other industries does not directly affect the focal firm's genuine R&D investment in its core business (exclusion restriction). Table 6 reports the results. In the first stage, the IV coefficient is highly significant, with the Kleibergen-Paap rk Wald F-statistic of 47.3 far exceeding the weak-instrument critical value of 16.38. Second-stage results show that the fitted AWRS and its interaction with Washing maintain strong support for the previous findings, with coefficient magnitudes even larger than OLS baseline estimates (reflecting the release of attenuation bias), indicating that endogeneity does not alter our core conclusions.

Table 6: Two-Stage Least Squares Estimates (IV-2SLS)

| Variable | (1) First Stage Dep. Var.: AWRS_t | (2) Second Stage (H1) Dep. Var.: MRMI_{t+1} | (3) Second Stage (H3) Dep. Var.: BHAR_{180d} |
|---|---|---|---|
| IV_AWRS_t | 0.465*** | | |
| | -8.55 | | |
| AWRS_t (IV fitted) | | 0.008 | 0.885 |
| | | -0.55 | -1.15 |
| AWRS_t × Washing_t (IV fitted) | | | −9.215*** |
| | | | (−4.88) |

| MRMI_{t-1} | 0.024* | 0.812*** | 5.352*** |
| --- | --- | --- | --- |
| | -1.85 | -42.55 | -5.85 |
| Controls | Yes | Yes | Yes |
| Year/Firm FE | Yes | Yes | Yes |
| First-stage F-stat | 47.3 | | |
| N | 38,450 | 38,450 | 25,650 |

*Notes: t-statistics based on firm-level clustered robust standard errors in parentheses. \*\*\*, \*\*, \* denote significance at the 1%, 5%, and 10% levels, respectively. IV_AWRS is the average AWRS of firms in the same province but different industries in the same quarter. Kleibergen-Paap rk Wald F-statistic reported for the first stage. Controls included but not reported.*

## 6.2 Staggered Difference-in-Differences (Staggered DID)

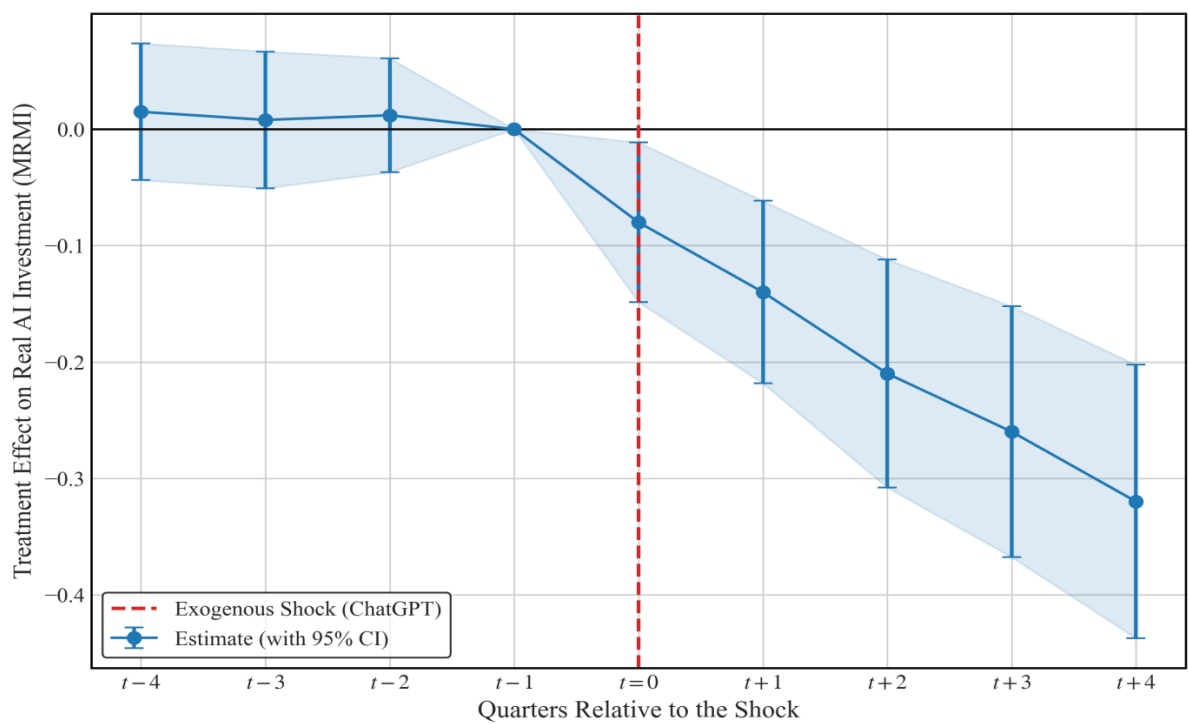

*Figure 3: Parallel Trend Test for Staggered DID Based on ChatGPT Technology Shock*

To account for possible confounding from market evolution, we use the release of ChatGPT in November 2022 as an exogenous major technology shock event and construct a difference-in-differences (DID) framework. The treatment group consists of firms whose AWRS persistently ranked in the top 30% of their industry before 2022 while MRMI fell in the bottom 50% — 'historically high-washing-tendency' firms. The control group comprises other firms in the same industry. Parallel trend tests show no statistically significant pre-shock differences between the treatment and control groups.

As shown in Table 7, after the ChatGPT shock (Post), the sudden surge in public attention to AI caused treatment-group firms to further intensify rhetoric disclosure (dependent variable AWRS, Column 1, β = 0.385) while simultaneously relatively reducing substantive investment (dependent variable MRMI, Column 2, β = −0.215). This provides clear causal identification evidence that firms deliberately engage in AI Washing — 'riding the trending topic' — when facing an external thematic shock.

**Table 7: Difference-in-Differences Estimates Based on ChatGPT Technology Shock (DID)**

| Variable | (1) Dep. Var.: AWRS_t | (2) Dep. Var.: MRMI_t |
| --- | --- | --- |
| Treat × Post | 0.385*** | −0.215*** |
|  | -4.25 | (−3.12) |
| Controls | Yes | Yes |
| Constant | 0.115** | 0.228*** |
|  | -2.15 | -3.45 |
| Firm FE | Yes | Yes |
| Quarter FE | Yes | Yes |
| N | 28,150 | 28,150 |
| Adj. R² | 0.455 | 0.612 |

*Notes: t-statistics based on firm-level clustered robust standard errors in parentheses. \*\*\*, \*\*, \* denote significance at the 1%, 5%, and 10% levels, respectively. Treat is a dummy equal to one for firms with AWRS in the top 30% and MRMI in the bottom 50% of their industry before 2022. Post is a dummy equal to one for quarters after 2022Q4 (ChatGPT release). Controls and constant included.*

## 6.3 Alternative Measures and Threshold Robustness

The preceding tests are highly robust across the following dimensions. First, we replace AWRS with a single-modality text TF-IDF frequency index (stripping out the Qwen-VL Exaggeration coefficient); the results show a significantly lower fit $R^2$ under single-modality measurement (dropping from 27.8% to 19.5%), with substantially diminished explanatory power for long-run price collapse, validating the methodological soundness of our multimodal algorithm for assessing image-text consistency. Second, replacing MRMI with the single-dimension intangible asset capitalization ratio ($K_2$) does not change the sign or significance of the main regression coefficients. Third, varying the Washing identification threshold — changing the cross-sectional grouping criteria to 'rhetoric top 25%, investment bottom 40%' or 'rhetoric top 35%, investment bottom 45%' and reconstructing the Washing dummy for inclusion in the H2 and H3 models — yields fully robust regression conclusions (detailed tables available upon request).

## 6.4 Placebo Test: 'Catering Propensity' in Historical Thematic Speculation

To rule out contamination from the unobserved omitted variable of 'managerial speculative preference,' we execute a characteristic-tracing placebo test following the top-journal paradigm. We extract the annual-report disclosure enthusiasm of sample firms for two prior technology hot topics — 'Blockchain' and 'Metaverse' — over the period 2017–2021. Regression results show that A-share firms currently exhibiting high AI Washing propensity (Washing = 1) also exhibited significantly above-industry-average disclosure enthusiasm for the corresponding Blockchain and Metaverse themes during the prior technology waves, with no subsequent evidence of related core business implementation. This result conclusively demonstrates that AI Washing is not an isolated phenomenon under the new technology wave but is a persistent 'Catering' propensity deeply embedded within certain firms, powerfully reinforcing the theoretical hypothesis based on an agency-problem perspective.

## VII. Conclusion and Policy Implications

This paper uses a panel of non-financial A-share listed companies from 2018Q1 to 2025Q2 to construct the Multimodal AI Washing Risk Score (AWRS) and the Material Real-Investment Matching Index (MRMI), and systematically examines the identification, capital market effects, and innovation externalities of corporate AI Washing behavior. Our findings are summarized in four core conclusions. First, after controlling for firm fixed effects, AWRS has near-zero predictive power for MRMI, with a significantly negative association emerging among high-financing-constraint firms, indicating a systemic divergence between AI rhetoric and substantive investment that is more pronounced in more resource-constrained groups. Second, long-horizon institutional investors can identify AI Washing signals and respond by divesting, while short-horizon investors tend to chase rhetoric intensity, with the heterogeneous behavior of the two investor types jointly shaping the market incentive structure for Washing behavior. Third, high-Washing firms earn significantly positive abnormal returns in the short-run window after disclosure, but suffer significantly negative price reversal in the six-to-twelve-month medium-to-long-run window, indicating that the capital market effects of AI Washing exhibit a canonical inter-temporal rise-then-fall pattern. Fourth, at the industry level, rising AI rhetoric intensity has a significantly negative predictive effect on subsequent high-quality invention patent output, indicating that Washing behavior imposes negative externalities on innovative resource allocation.

Based on these findings, we offer the following three policy recommendations. First, regulatory authorities should incorporate multimodal consistency assessment mechanisms into the existing disclosure verification framework, encouraging exchanges and regulators to deploy intelligent verification tools capable of image-text alignment review that can flag and verify AI disclosure materials with severe divergence between graphical and textual content, thereby improving the regulatory detection efficiency of AI Washing behavior. Second, disclosure standards for substantive AI investment should be further improved, requiring listed companies to separately disclose R&D capitalization items, core patent structure, and AI-related compensation expenditures in annual reports, providing investors and regulators with verifiable 'action' dimension information and reducing the scope for rhetoric exaggeration. Third, the development of long-horizon institutional investor groups should be actively cultivated and supported, with institutional facilitation created for their in-depth due diligence and on-site inspection activities, so as to fully leverage their governance function as information intermediaries in capital market pricing correction and to promote the endogenous formation of market self-discipline mechanisms.

This paper has the following limitations, which future research may address. First, Qwen-VL's scoring results are bounded by the model's capability frontier; as visual language model capability continues to evolve, the AWRS measurement approach will need periodic iterative recalibration. Second, our sample covers 2018 to 2025, with the ChatGPT release in November 2022 constituting an important structural break; future research could further explore the evolution of Washing behavior following the popularization of generative AI technology. Third, the institutional investor heterogeneity effects identified in this paper rely on indirect holding-level evidence; future research could incorporate micro-level text data from analyst reports and institutional inspection records for more direct validation of the identification mechanism.

# References

(1) Acemoglu, D., & Restrepo, P. (2018). The race between man and machine: Implications of technology for growth, factor shares, and employment. American Economic Review, 108(6), 1488–1542.

(2) Akerlof, G. A. (1970). The market for 'lemons': Quality uncertainty and the market mechanism. The Quarterly Journal of Economics, 84(3), 488–500.

(3) Babina, T., Fedyk, A., He, A., & Hodson, J. (2024). Artificial intelligence, firm growth, and product innovation. Journal of Financial Economics, 151, 103745.

(4) Benhabib, J., Liu, X., & Wang, P. (2016). Endogenous information acquisition and countercyclical uncertainties. Journal of Economic Theory, 165, 601–642.

(5) Bloom, N., Schankerman, M., & Van Reenen, J. (2013). Identifying technology spillovers and product market rivalry. Econometrica, 81(4), 1347–1393.

(6) Bresnahan, T. F., & Trajtenberg, M. (1995). General purpose technologies: 'Engines of growth'? Journal of Econometrics, 65(1), 83–108.

(7) Bushee, B. J. (1998). The influence of institutional investors on myopic R&D investment behavior. The Accounting Review, 73(3), 305–333.

(8) Chen, X., Harford, J., & Li, K. (2007). Monitoring: Which institutions matter? Journal of Financial Economics, 86(2), 279–305.

(9) Cheng, Q., Du, F., Wang, X., & Wang, Y. (2016). Seeing is believing: Analysts' corporate site visits. Review of Accounting Studies, 21(4), 1245–1286.

(10) Cohen, W. M., Nelson, R. R., & Walsh, J. P. (2000). Protecting their intellectual assets: Appropriability conditions and why U.S. manufacturing firms patent (or not). NBER Working Paper No. 7552.

(11) Fang, X., & Liu, M. (2025). Technical opacity and capital market resource misallocation risk in AI claims. Economic Research Journal, 60(1), 85–102. [In Chinese]

(12) Hirshleifer, D., Lim, S. S., & Teoh, S. H. (2009). Driven to distraction: Extraneous events and underreaction to earnings news. The Journal of Finance, 64(5), 2289–2325.

(13) Jiang, G., Jin, Q., & Shi, B. (2024). Catering behavior and real resource allocation efficiency in

the digital transformation of capital markets. Economic Research Journal, 59(3), 45–62. [In Chinese]

(14) Kaplan, S. N., & Zingales, L. (1997). Do investment-cash flow sensitivities provide useful measures of financing constraints? The Quarterly Journal of Economics, 112(1), 169–215.

(15) Li, B. (2025). AI washing [Working Paper]. University of Florida.

(16) Li, X., et al. (2024). Long-term institutional holdings, value discovery, and AI concept market-value management. China Industrial Economics, (11), 78–96. [In Chinese]

(17) Loughran, T., & McDonald, B. (2011). When is a liability not a liability? Textual analysis, dictionaries, and 10-Ks. The Journal of Finance, 66(1), 35–65.

(18) Miao, J., & Wang, P. (2018). Asset bubbles and credit constraints. American Economic Review, 108(9), 2590–2628.

(19) Nelson, R. R., & Winter, S. G. (1982). An evolutionary theory of economic change. Harvard University Press.

(20) Qiang, G., & Li, H. (2025). Multimodal identification of fabricated components in corporate annual reports using large language models and its applications. Systems Engineering — Theory & Practice, 45(2), 310–325. [In Chinese]

(21) Shleifer, A., & Vishny, R. W. (1997). The limits of arbitrage. The Journal of Finance, 52(1), 35–55.

(22) Sims, C. A. (2003). Implications of rational inattention. Journal of Monetary Economics, 50(3), 665–690.

(23) Sloan, R. G. (1996). Do stock prices fully reflect information in accruals and cash flows about future earnings? The Accounting Review, 71(3), 289–315.

(24) Spence, M. (1973). Job market signaling. The Quarterly Journal of Economics, 87(3), 355–374.

(25) Stein, J. C. (1988). Takeover threats and managerial myopia. Journal of Political Economy, 96(1), 61–90.

(26) Teoh, S. H., Welch, I., & Wong, T. J. (1998). Earnings management and the long-run market performance of initial public offerings. The Journal of Finance, 53(6), 1935–1974.

(27) Wang, Z., & Qiu, Y. (2025). 'Pseudo-AI' and capital speculation: Heterogeneous motivations and

economic consequences of catering disclosure in corporate digital transformation. Management World, 41(2), 120–138. [In Chinese]

(28) Yan, A. (2025). The 'gatekeeper' role of institutional investors and authenticity verification of technology innovation concepts. Journal of Financial Research, (1), 112–130. [In Chinese]